\documentstyle[epsfig,rotate]{elsart}

\begin{document}

\newcommand{\beq}{\begin{equation}}
\newcommand{\eeq}{\end{equation}}
\newcommand{\feh}{\hbox{$[{\rm Fe}/{\rm H}]$}}
\newcommand{\mvrr}{\hbox {$\rm M_v(RR)$}}
\newcommand{\dvtwo}{\Delta {\rm V}}
\newcommand{\dv}{\hbox {$\Delta \rm V^{TO}_{HB}$}}

\rightline{{CWRU-P27-99}}

\begin{frontmatter}
\title{The Age of Globular Clusters}
\author[ad1]{Lawrence M. Krauss \thanksref{corr}},
\thanks[corr]{E-mail: krauss@theory1.phys.cwru.edu}
\address[ad1]{Departments of Physics and Astronomy, Case Western Reserve 
University, 10900 Euclid Ave. \cty
Cleveland OH,
\cny USA}

\begin{abstract}

I review here recent developments which have
affected our understanding of both the absolute age of globular clusters and the
uncertainties in this age estimate, and comment on the implications for
cosmological models. This present estimate is in agreement with the range long
advocated by David Schramm.  The major uncertainty in determining ages of
globular clusers based upon the absolute magnitude of the main sequence
turn-off remains the uncertainty in the distance to these clusters. 
Estimates of these distances have recently been upwardly revised due
to Hipparcos parallax measurements, if one calibrates luminosities of
main sequence stars.  However, it is
important to realize that at the present time, different distance measures are
in disagreement.  A recent estimate is that the oldest clusters are $11.5
\pm 1.3$ Gyr, implying a one-sided $95\%$ confidence level lower limit of 9.5
Gyr, if statistical parallax distance measures are not incorporated. 
Incorporating more recent measures, including Hipparcos based 
statistical parallax measures, raises the mean predicted age to $12.8 \pm 1$
Gyr,
with a 95 $\%$ confidence range of 10-17 Gyr.  I conclude by discussing
possible improvements which may allow a more precise age distribution in
the near future.
\end{abstract}

\end{frontmatter}

\section{Introduction and Overview: Some Personal Reflections}
I remember the first time I discussed the globular cluster age problem with
David Schramm.  This was long before I knew much at all about the detailed
issues associated with fitting the main sequence turn-off magnitudes.  At the
time I was at Yale University, and my colleague in the Astronomy Department
there, Pierre Demarque, was using the new Yale isochrones and finding good
agreement with ages for the oldest globular clusters in the range 16-20 Gyr. 
Pierre suggested that this number was accurate to perhaps 20 $\%$, although he
felt that this could just as likely result in
 longer ages rather than shorter ones.    This age estimate was clearly in
conflict with the estimate for the age of a flat matter-dominated Universe,
then the preferred cosmological model, $t_{universe} = 6.6 (1/h)$ Gyr, unless
the Hubble constant $H_0 =100 \ h \ km s^{-1} Mpc^{-1} $ was uncomfortably
small.  

When I spoke to David about this apparent further confirmation of the
long-standing age problem, he smiled and with his usual confidence he  asserted
(colored of course by his firm belief in a flat Universe, as predicted by
Inflationary models)  that there were likely to be additional systematic
uncertainties which could shift the allowed range.  In the end he felt the
allowed age range would be closer to 10-14 Gyr.   I was somewhat surprised at
the time by his confidence in this claim, but I shouldn't have been.  David had
an astute sense of what the key issues were in astrophysics, and where there
were weaknesses or loopholes, even if he didn't always annunciate these in
public.

At around that time I was investigating another issue of great interest to
David, Big Bang Nucleosynthesis.  I had decided to utilize Monte
Carlo techniques to determine the actual theoretical uncertainties in BBN
predictions for light element abundances.  At that time computational
resources had advanced to the point that it was practical to alter BBN codes to
run many different times with individual nuclear reaction rates chosen at
random from various distributions with ranges appropriate to the individual
experimental uncertainties.  This allowed one to quote quantitative confidence
limits on BBN predictions, and also to explicitly explore the dominant
uncertainties in the analysis.  

Almost a decade later, after moving to Case Western Reserve University, I
decided to take David's concerns about globular cluster age estimates to heart,
and attempt a similar analysis in this regard.   I contacted Pierre Demarque,
and his former student Brian Chaboyer, who I also knew from Yale, and who was
then a postdoc at CITA in Toronto.  Brian and Pierre not only had good stellar
evolution codes, but they were fully familiar with the important observational
literature, which we would have to scour in order to assess the input
uncertainties in the globular cluster age estimates, and equally important, to
assess the fits to observations.  Peter Kernan at CWRU and I had familiarity,
from our BBN work, with Monte Carlo techniques and the related statistical
analysis of data, and so it seemed like a good combination.  Our geographic
proximity allowed us to meet together to go over each facet of the input data
in order to agree on appropriate uncertainties, and then we were able to
rewrite the stellar evolution codes to accomodate a Monte Carlo over the
following months.  Moreover, because of another quantum leap in
computational resources, one could now run a stellar evolution code to
produce a set of isochrones in several minutes, so it was feasible to
sample millions of different models using several months of dedicated
computer time.

Our first analysis\cite{chab1} suggested a best fit median age of 14.2
Gyr, and several other groups at the time also reported best fit ages in
the 14-15 Gyr range, based on independent methods, and differing input
physics.  However, it also appeared that existing uncertainties could
allow, at the 95$\%$ confidence level, ages as low as 11.8 Gyr.  This was
still somewhat uncomfortable for a flat matter- dominated Universe, given
the Hubble Key Project estimate of
$H_0
\approx 80$ for the Hubble constant, but the disagreement was much smaller, and
David was quite enthused by the results.  Our other chief result confirmed
that it was not stellar model uncertainties which dominated the overall
uncertainty in our globular cluster age estimates, but rather the
observational uncertainty in the distance to globular clusters.  Because
we normalized our absolute magnitude to the Horizontal Branch RR Lyrae
stars, this distance uncertainty translated into an uncertainty in the
the RR Lyrae distance modulus.

Then about a year after these analyses,
the Hipparcos satellite produced its catalogue of parallaxes of nearby stars,
causing an apparent revision in distance estimates.  The Hipparcos parallaxes
seemed to be systematically smaller, for the smallest measured parallaxes, than
previous terrestrially determined parallaxes.  Could this represent the
unanticipated systematic uncertainty that David has suspected?  Since all the
detailed analyses had been pre-Hipparcos, several groups scrambled to
incorporate the Hipparcos catalogue into their analyses.  The immediate result
was a generally lower mean age estimate, reducing the mean value to 11.5-12 Gyr,
and allowing ages of the oldest globular clusters as low as 9.5 Gyr.   However,
what is also clear is that there is now an explicit systematic uncertainty in
the RR Lyrae distance modulus which dominates the results.  Different
measurements are no longer consistent.  Depending upon which distance estimator
is correct, and there is now better evidence that the distance estimators which
disagree with Hipparcos-based main sequence fitting should not be dismissed out
of hand, the best-fit globular cluster estimate could shift up perhaps $1
\sigma$, or about 1.5 Gyr, to about 13 Gyr.

While all this has happened, a number of other important revolutions have been
taking place in observational cosmology.   The HST Key Project has lowered
their best fit Hubble Constant value to $H_0 =70 \pm 7$ \cite{Freeman},
raising the upper limit on the allowed age of the Universe for a given
cosmological model.  At the same time, observations of Type 1a Supernovae
have provided direct evidence in support of the growing suspicion that the
cosmological constant is non-zero.  Previously the cosmological constant
was invoked as one way out of the age problem, as it can raise the age of
a flat Universe by an arbitrary amount for a fixed Hubble constant,
depending upon the value of the cosmological constant.  If the
cosmological constant is indeed non-zero, then one will have no
difficulty reconciling globular cluster ages with the Hubble age.  If it
turns out to be zero, we cannot yet definitively rule out a flat matter
dominated universe on the basis of globular cluster ages alone, although
the current results require pushing all uncertainties to their limit in
order to get concordance.

It is a pleasure to dedicate this personal overview of recent developments in
Globular Cluster age estimation to David's memory. 
He helped inspired my own interest in trying to pin down globular cluster ages,
and it is satisfying that the results seem to at least partly confirm his own
suspicions.   It also goes without saying that much of what I will describe
here I learned from my collaborators.

\section{Main Sequence Fitting of Globular Cluster Ages: An Overview}

This will not be an encyclopedic overview.  There are many good reviews of the
field \cite{reviews,chab1}.   I will try and stress the key features that underlie
different estimates, and which have been affected by recent developments.

The basic idea behind main sequence fitting is simple.  A stellar model is
constructed by solving the basic equations of stellar structure, including
conservation of mass and energy and the assumption of hydrostatic equilibrium,
and the equations of energy transport.  Boundary conditions at the center
of the star and at the surface are then used, and combined with assumed
equation of state equations, opacities, and nuclear reaction rates in
order to evolve a star of given mass, and elemental composition.

Globular clusters are compact stellar systems containing up to $10^5$ stars,
with low heavy element abundance.  Many are located in a spherical halo around
the galactic center, suggesting they formed early in the history of our
galaxy.  By making a cut on those clusters with large halo velocities, and
lowest metallicities (less than 1/100th the solar value), one attempts to
observationally distinguish the oldest such systems. Because these systems are
compact, one can safely assume that all the stars within them formed at
approximately the same time.

Observers measure the color and luminosity of stars in such clusters, producing
color-magnitude diagrams of the type shown in Figure 1.

\begin{figure}[tbp]
  \leavevmode\center{\epsfig{figure=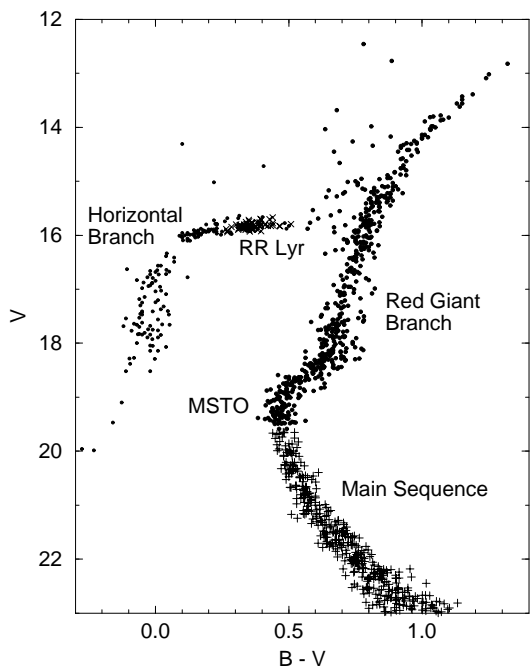,width=10cm}}
\caption{Color-magnitude diagram for a typical globular cluster,
M15\cite{durr}. Vertical axis plots the magnitude (luminosity) of the
stars in the V wavelength region and the horizontal axis plots the color
(surface temperature) of the stars.}
\label{fig:CMDIAG}
\end{figure}

Next, using stellar models, one can attempt to evolve stars of differing
mass for the metallicities appropriate to a given cluster, in order to fit
observations.  A point which is often conveniently chosen is the so-called main
sequence-turnoff (MSTO) point, the point in which hydrogen burning (main
sequence) stars have exhausted their supply of hydrogen in the core.  After the
MSTO, the stars quickly expand, become brighter, and are referred to as Red
Giant Branch (RGB) stars.  Higher mass stars develop a helium core that is so
hot and dense that helium fusion begins.  These form along the horizontal
branch.  Some stars along this branch are unstable to radial pulsations, the
so-called RR Lyrae stars mentioned earlier, which are important distance
indicators.  While one in principle could attempt to fit theoretical isochrones
(the locus of points on the predicted CM curve corresponding to different mass
stars which have evolved to a specified age), to observations at any point, the
main sequence turnoff is both sensitive to age, and involves minimal
(though just how minimal remains to be seen) theoretical uncertainties.

Dimensional analysis tells us that the main sequence turnoff should be a
sensitive function of age.  The luminosity of main sequence stars is very
roughly proportional to the third power of solar mass.  Hence the time it takes
to burn the hydrogen fuel is proportional to the total amount of fuel
(proportional to the mass M), divided by the Luminosity---
proportional to $M^3$.  Hence the lifetime of stars on the main sequence
is roughly proportional to the inverse square of the stellar mass.

Of course the ability to go beyond this rough approximation depends completely
on the on the confidence one has in one's stellar models.  It is worth
noting that several improvements in stellar modeling have recently
combined to lower the overall age estimates of globular clusters.  The
inclusion of diffusion lowers the age of globular clusters by about 7$\%$
\cite{eighteen}, and a recently improved equation of state which incorporates
the effect of Coulomb interactions \cite{nineteen} has lead to a further
7$\%$ reduction in overall ages.   Of course, what is most important for
the comparison of cosmological predictions with inferred age estimates is
the uncertainties in these and other stellar model parameters, and
not merely their best fit values. 

The uncertainties in determining each of these parameters leads to
uncertainties in fitting the age of globular clusters.  One of the advantages
of determining globular cluster ages by fitting the MSTO is that the low
metallicity main sequence stellar models are relatively simple, so that some of
the theoretical complexities of solar physics that plague attempts to understand
certain classes of stars are minimized here.  In particular, probably the least
understood aspect of stellar modeling involves the treatment of convection. 
Main sequence and red giant stars have surface convection, so that the surface
properties such as color are rather uncertain, whereas horizontal branch stars
have convective cores and thus the predicted luminosities and lifetimes of
these stars are highly uncertain.  

The remaining key
parameter uncertainties of these main sequence stellar models
include: pp and CNO chain nuclear reaction rates, stellar opacity
uncertainties, mixing length, diffusion uncertainties, helium abundance
uncertainties, and uncertainties in the abundance of the $\alpha$-capture
elements (O, Mg, Si, S, and Ca).

\section{Monte Carlo Estimates of Age Uncertainties resulting from Model
Parameter Uncertainties}

In order to account for the impact of these uncertainties in the input
parameters on the eventual derived ages, one can take a Monte Carlo approach.
In this case, many different stellar models are run on a computer.  In each
model different values of the input parameters are chosen, and these values are
allowed to run over a distribution which is based on the assumed uncertainty in
each parameter.  If the uncertainty is dominated by statistics, a gaussian
distribution in this variable is chosen.  If systematics dominate, as is often
the case, a top hat distribution is usually chosen \cite{chab1,chab2}. 

The set of input parameters, and the range chosen for the
figures displayed here is given in table 1.

\begin{table}\caption{Monte Carlo Stellar Model Input Parameters}
\begin{tabular}{|l|l|l|} \hline
Parameter & Distribution & Comment \\ \hline 
mixing length & $1.85\pm 0.25$ (stat.) & fits GC observations \\
helium diffusion coefficients & 0.3 -- 1.2 (syst.) &
sys. error  dominate \\
high temperature opacities & $1\pm 0.01$ (stat.) & comparison of
OPAL \\
~~ & & \& LAOL opacities \\
low temperature opacities & $0.7 - 1.3$ (syst.) & comparing  
different tables \\
primordial $^4$He abundance & $0.22 - 0.25$ (syst.)&
sys. error dominate \\
oxygen abundance, [O/Fe] & $+0.55\pm 0.05$ (stat.) & mean from  
\cite{nissen} \\
& $\pm 0.20$(syst.) &\\
surface boundary condition & &{grey or \cite{kris}} \\
colour table & &{\cite{ryi} or \cite{kurcol}}
 \\
Nuclear Reaction Rates: & & \\
$p + p \longrightarrow {^2\rm H} + e^+ + \nu_e$ & $1\pm 0.002$ (stat.)
& see \cite{chab1} \\
& $^{+0.0014}_{-0.0009}\,\,^{+0.02}_{-0.012}$ (syst.)  & \\
${\rm {^3He} + {^3He} \longrightarrow {^4He}} + 2p$ & $1\pm 0.06$  
(stat.)&
\cite{bahcpin}  \\
${\rm {^3He} + {^4He} \longrightarrow {^7Be}} + \gamma$ & $1\pm 0.032$  
(stat.)
& \cite{bahcpin} \\
${\rm {^{12}C} + p \longrightarrow {^{13}N}} + \gamma$ & $1\pm 0.15$  
(stat.) 

& \cite{bahcall} \\
${\rm {^{13}C} + p \longrightarrow {^{14}N}} + \gamma$ & $1\pm 0.15$  
(stat.) 

& \cite{bahcall} \\
${\rm {^{14}N} + p \longrightarrow {^{15}O}} + \gamma$ & $1\pm 0.12$  
(stat.)
& \cite{bahcall} \\
${\rm {^{16}O} + p \longrightarrow {^{17}F}} + \gamma$ & $1\pm 0.16$  
(stat.)
& \cite{bahcall} \\ \hline
\end{tabular} 
\end{table}

When this analysis is completed, one can explore the sensitivity of inferred
ages to individual input parameters by plotting this age as a function of the
chosen parameter for each stellar model run.  An analytical fit to the
determined age, as a function of the relevant input parameter  can then be
derived.  It turns out  that the dominant uncertainty in theoretical
models is due to the uncertainty in the abundance of $\alpha$-capture
elements (with oxygen being the dominant such element).  Estimates of the
oxygen abundance, for example, vary by up to a factor of 3.

As an example of the sensitivity of inferred ages to variations in input
parameters I display in Figure 2, the inferred age as a function of the assumed
logarithmic abundance of $\alpha$-capture elements relative to iron
\cite{chab2}.  The best fit median along with $\pm 1 \sigma$ limits is also
plotted.  These lines are of the form $t_9 = a + b[\alpha/{\rm Fe}]$ with the
following coefficients: median $(a,b) =(13.83, -3.77)$, $-1 \sigma (a,b)
=(13.26,-3.72)$,  $+1 \sigma (a,b) =(14.54,-4.00)$.

\begin{figure}[tbp]
  \leavevmode\center{\rotate[r]{\epsfig{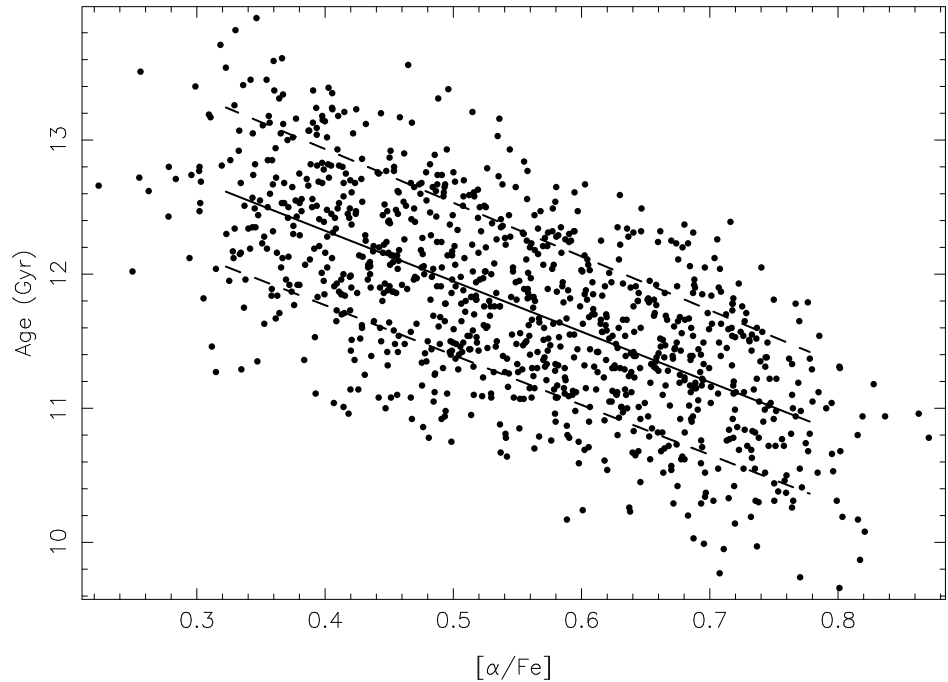}}}
\caption{Sensitivity of inferred globular cluster ages to abundance of 
$\alpha$-capture elements relative to iron}
\label{fig:alpha}
\end{figure}

Similar curves can be derived for the sensitivity of MSTO age estimates to the
other stellar physics input parameters \cite{chab2}.  The net effect of all
such uncertainties is in any case relatively small, at the level of $7\%$ or
less.   

\section{Observational Uncertainties and Globular Cluster Ages}

It turns out, however, that the dominant uncertainty in the use
of the MSTO luminosity for determining the age of globular clusters arises
from the comparison of theoretical predictions to observations.  In
particular, normalizing the predicted luminosity curves to observed
magnitudes requires a distance measurement to the cluster.  Moreover,
because of uncertainties in the effective surface temperatures of the
models, and to remove sensitivity to reddening \cite{reviews} 
the turnoff luminosity is compared to the Horizontal branch luminosity as
an age discriminant.  Specifically, one considers the difference in
magnitude between the HB and the MSTO, $\Delta V^{TO}_{HB}$.  
Furthermore, since the theoretically determined HB luminosity is subject
to large uncertainties due to convective effects in the core, one
utilizes the observed HB luminosities and the theoretical MSTO
luminosities in this subtraction.   Determining the absolute luminosity
of the HB branch revolves around determining the distance to the
cluster.   One can parametrize the uncertainty in this distance
determination by the uncertainty in the empirical calibration of the
absolute magnitude,
$M_v(\rm RR)$, of RR Lyrae stars located on the HB.

With a calibration of $M_v(\rm RR)$, one can then use theoretically
derived values for $M_v(\rm TO)$ to determine a grid
of predicted $\Delta V^{TO}_{HB}$ values as a function of age and [Fe/H]
that is fit to an equation of the form

\begin{equation}
t_9 = \beta_0 + \beta_1\dvtwo + \beta_2\dvtwo^2 + \beta_3\feh +
\beta_4\feh^2 + \beta_5\dvtwo\feh,
\label{fit}
\end{equation}
The observed values of \dv ~and \feh, 
along with their corresponding errors, are then input in
(\ref{fit}) to determine the age and its error for each GC in the
sample.

\begin{figure}[tbp]
  \leavevmode\center{\rotate[r]{\epsfig{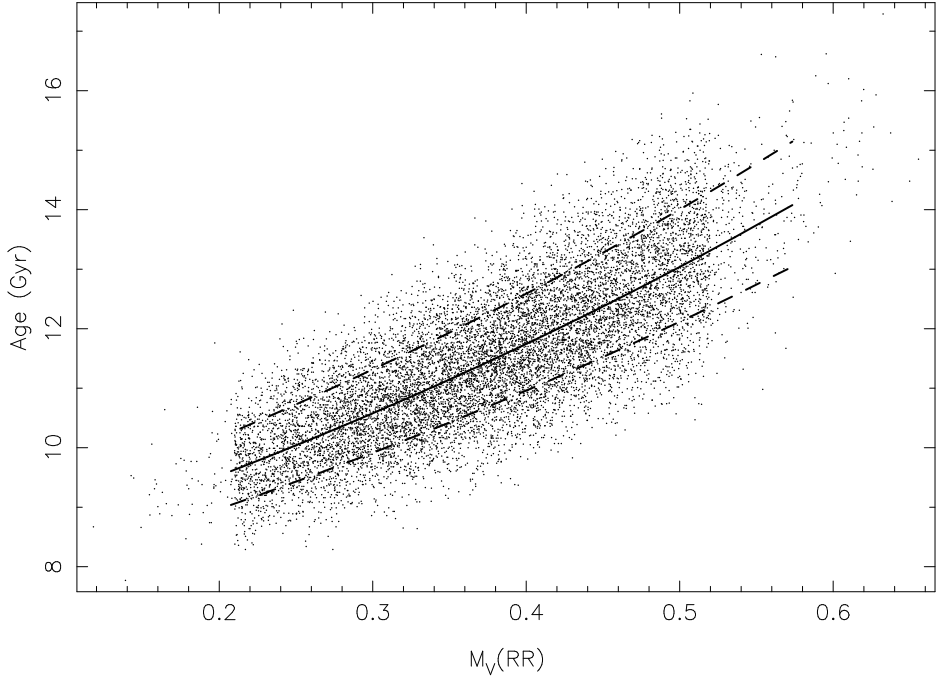}}}
\caption{Sensitivity of inferred globular cluster ages to the adopted
value of $M_v(\rm RR)$}
\label{fig:mvrr}
\end{figure}

The uncertainty in determining $M_v(\rm RR)$, and hence the distance to
globular clusters far outweighs any intrinsic stellar model
uncertainties as far as globular cluster age determinations are
concerned.  If we normalize
$M_v(\rm RR)$ at [Fe/H]
$=-1.9$, then we can display the dependence of globular cluster ages from
the fitting equation (\ref{fit}) on this value of $M_v(\rm RR)$ for the
locus of evolved stars resulting from the Monte Carlo analysis.  This is
shown in Figure 3.

\section{Distance Estimators and the Age of the Oldest Globular Clusters}

In order to derive a reliable age range, then, one must examine in some
detail the uncertainties in distance determinations to globular
clusters.  There are five independent distance determinators that we have
used in our own analyses.  As described above, with the \dv 
age-determination  technique these can all be translated into an
uncertainty in $M_v(\rm RR)$. Because the various determinations are
appropriate for systems of differing metallicity, one must translate the
constraints on $M_v(\rm RR)$ to the metallicities appropriate to globular
clusters.  This of course also  introduces uncertainties into the
analysis.  One assumes a simple linear relation, which we normalize at
the mean metallicity of interest:
\begin{equation}
M_v(\rm RR) =\alpha [{\rm Fe/H} + 1.9] +  \beta
\end{equation}
Various deteminations of this relation imply a weighted mean slope
$\alpha = .23 \pm .04$ \cite{chab1,alpha}, which we have used to translate the
distance  measures to luminosity determinations at [Fe/H]$=-1.9$.

I briefly describe the different distance determinators, and how we
combined them in our analyses of globular cluster ages, and compare
these with independent analyses by other groups.  In the next section, I
will discuss more recent results and their effect on age determinations.

\subsection{Statistical Parallax:}
This is a traditional method used to determine RR Lyrae star absolute
magnitude, and has been applied to stars in the field \cite{Layden}.  
It involves the
use of observed proper motions and radial velocities 
with statistical estimates for the total velocities, in order to 
infer distances to these systems.  Because this yields estimates for
$M_v(\rm RR)$ for stars in the field, we did not, in our 1998 analysis, 
include this distance measure, as we reasoned that there could be systematic
effects differentiating this sample from globular cluster RR Lyrae stars.
As I shall describe, there are more recent reasons to believe that this is
not the case.  It is also interesting that the Hipparcos satellite provided
a large number of proper motions which can be used in this analysis, the net
effect of which is to drive $M_v(\rm RR)$ in the {\it opposite} direction from
the direction favored by main sequence fitting due to Hipparcos, as I 
discuss shortly.

\subsection{Astrometric Distances} This method is in fact similar to the
statistical parallax method, except that it is applied directly to
globular clusters \cite{c}.  
It requires a dynamical model for the cluster, but
nevertheless provides an otherwise direct measure of distance, independent
of a ``standard candle''.

\subsection{LMC Distance Measures}
Several independent distance measures to the LMC allow the calibration of
the magnitude of RR Lyrae stars in this system.  These distance measures
include the use of Cepheid variable period-luminosity relations 
\cite{walker,madore},
or SN 1987A light echoes \cite{gould,sonn}.

\subsection{Theoretical HB Models}
As indicated earlier, theoretical modeling of the luminosity of 
HB stars is subject to possible large
systematic uncertainties due to the effects of convection in the cores of
these systems, and on great sensitivity to the assumed primordial
helium abundance. Nevertheless, theoretical models have improved greatly 
\cite{broc,others},
and theoretical predictions can be compared with observations to attempt
to constrain distances to globular clusters using this method.

\subsection{Main Sequence Fitting}
If one can measure parallaxes to nearby field stars, one can
attempt to determine the position of the zero age main sequence, and then,
via comparison to globular cluster CM diagrams, obtain a direct estimate to
the distance to the cluster.  Indeed, 
perhaps the greatest shift in estimates of globular cluster ages came
about as a result of the parallax data from the Hipparcos satellite. 
Hipparcos parallaxes appeared to be systematically smaller than those
obtained from the ground, suggesting brighter stars, which in turn would
suggest shorter times for evolution off the main sequence.  However, there
are several problems with a naive extrapolation of the Hipparcos results
to globular cluster age estimates.  In particular, the Hipparcos stars
are not measured at similar metallicities in most cases, and metallicity-
dependence could easily introduce uncertainties which can overwhelm any
systematic shifts due to the smaller parallaxes.

In any case, the excitement over the Hipparcos results has caused a 
number of groups to determine distances to globular clusters on the
basis of main sequence fitting \cite{reid,pont,gratton,chab2}.  
Care must be taken to
isolate those stars with low metallicity, and also to remove possible
binary systems from the analysis \cite{chab2}.

\subsection{White Dwarf Fitting}
A similar analysis can compare the parallaxes of local white dwarfs
to the recent accurate photometric measurements from the HST of white
dwarfs in clusters to dtermine distances \cite{renzinietal}.

\subsection{Combined Results}

One can combine the results of the different measures above, along with
an attempt to estimate statistical and systematic uncertainties in each
method, in order to determine an appropriate value of $M_v(\rm RR)$ for
use in globular cluster age estimates.  I display the individual results
in Table 2.

\begin{table}[h]
\caption{\mvrr\ Calibration}
\begin{tabular}{|l|l|l| c|}
\hline
Method & \feh & \mvrr & \mvrr at $\feh = -1.9$ \\ \hline
Statistical Parallax & $-1.6$ & $0.77 \pm 0.13$ & $0.70 \pm 0.13$ \\
Astrometric & $-1.59$ & $0.59\pm 0.11$ & $0.52\pm 0.11$ \\
White dwarf fitting to N6752 & $-1.51$ & $0.45\pm 0.14$ & $0.36\pm  
0.14$ \\
Subdwarf fitting to N6752 & $-1.51$ & $0.30\pm 0.15$ & $0.21\pm 0.15$  
\\
Subdwarf fitting to M5 & $-1.17$ & $0.54\pm 0.09$ & $0.37\pm 0.09$\\
Subdwarf fitting to M13 & $-1.58$ & $0.36\pm 0.14$ & $0.29\pm 0.14$\\
LMC RR Lyr & $-1.90$ &$0.44\pm 0.14$ &$0.44\pm  
0.14$\\
Theoretical HB models  & $-2.20$ & $0.36\pm 0.10$ & $0.43\pm 0.10$\\ \hline
\end{tabular}
\end{table}

Clearly, the final range of ages determined for globular clusters will depend
on how one combines these different, often inconsistent,  
distance determinators.   In our
comprehensive analysis of 1998, we did not include the statistical parallax
results, and combined the other results using a top-hat distribution which 
ranged from the lowest mean value to the highest.  Sampling randomly over
such a distribution, and incorporating the Monte Carlo age estimates for
the 4 million stellar models we evolved, we determined a best estimate
age for the 17 oldest globular clusters to be $11.5 \pm 1.3$ Gyr, with a
one-sided $95\%$ confidence lower limit of 9.5 Gyr.   The derived 
distribution\cite{chab2} is
shown in Figure 4.

\begin{figure}[tbp]
  \leavevmode\center{\rotate[r]{\epsfig{figure=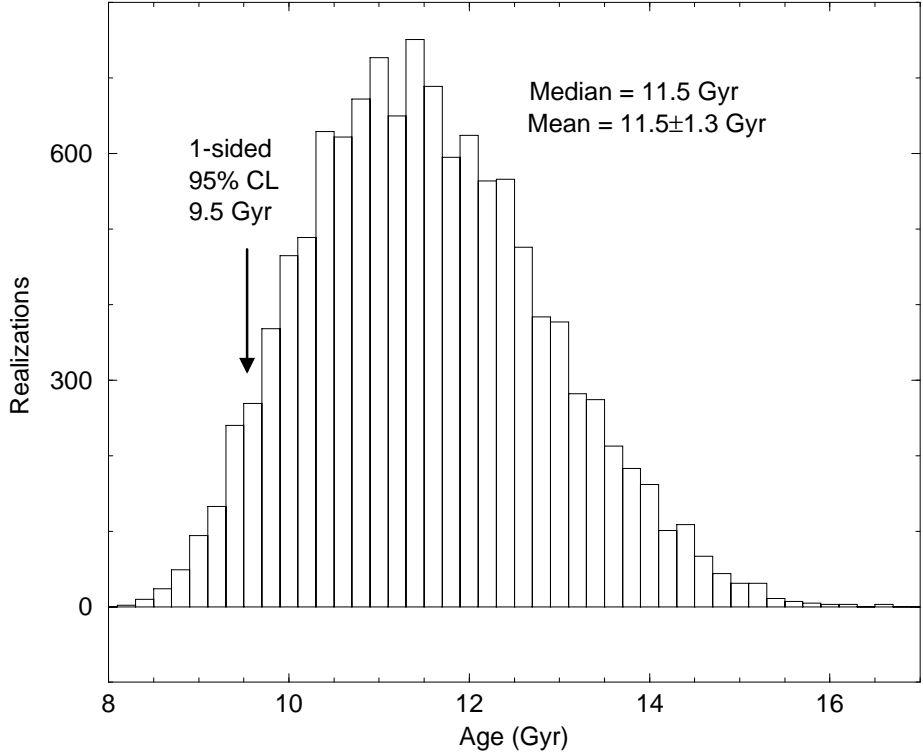,width=12cm}}}
\caption{Histogram of globular cluster ages}
\label{fig:globclust}
\end{figure}

It is heartening, perhaps, that several independent analyses, all using
main sequence turn-off as an age indicator, but using different stellar
models and input physics, and different globular cluster distance
estimators, nevertheless produce roughly consistent age estimates (see
Table 3).  All
of these are roughly 2-3 Gyr younger than previous estimates, due in
part to the new longer distance scale suggested by Hipparcos, and also due
to improved input physics in the models. 

\begin{table}[h]
\caption{Different Main Sequence Turn-off Estimate for Globular Cluster Ages}
\begin{tabular}{|l|l|c|}
\hline
\hline
Age (Gyr) & Distance determinator & Study \\ \hline
$11.5 \pm 1.3$ & 5 techniques (discussed here) & \cite{chab2} \\
$12 \pm 1$ & main sequence fitting (post Hipparcos) & \cite{reid} \\
$11.8 \pm 1.2$ & main sequence fitting (post Hipparcos) & \cite{gratton} \\
$12 \pm 1$ & Theoretical HB models, main sequence fitting (post H.)
  & \cite{danton} \\
$12.2 \pm 1.8$ & Theoretical HB models  & \cite{sal} \\ \hline
\end{tabular}
\end{table}

\section{Conclusions: Work in Progress}

Where do we go from here?  The results of recent years suggest younger
ages than previously determined for globular clusters, and also give us
more reliable estimates of the uncertainties in the presumed ages, just
as David would have liked.  
Nevertheless, the current results, while barely marginally consistent
with a flat matter dominated Universe for the recent HST Key Project
Hubble constant measurement, are still consistent with a vast range
of cosmological models.  It is clear that systematic uncertainties
dominate the analysis, and little significant improvement will be 
possible until consistency is achieved between different age estimators.

There have nevertheless been several recent observational results that
suggest a possible shift from the mean globular cluster age estimate
quoted above. In particular, a recent comparison of Hipparcos main
sequence fitting  for subdwarfs with LMC distance measures has suggested
that this Hipparcos distance estimator is perhaps $1 \sigma$ too
long
\cite{gratt99}, leading to a revised age estimate of approximately 13
Gyr.  Also, recent studies of the field RR Lyrae stars used in
statistical parallax studies suggest that previous concerns about
systematic shifts between field and cluster stars  may have been
unwarranted.  We have recently completed a preliminary analysis of
globular cluster ages, including the statistical parallax
measure\cite{Layd99},  and also updating various stellar
model parameters \cite{chabkrauss}, and arrive at a new mean value
of $12.8 \pm 1$ Gyr, with a 95 $\%$ confidence range of 10-17 Gyr.  This 
age range is now inconsistent with a flat matter dominated universe for
the HST Key project value for $H_0$. 

At the same time as these developments have taken place, important new
developments have occurred in our understanding of BBN, a subject near
and dear to David's heart.  New observations of primordial deuterium suggest
a primordial helium abundance of perhaps 24.5-25 $\%$  If this the case,
then the best fit age may be lowered by perhaps .3 Gyr.  While it will not
significantly alter the conclusions one might draw, I find it poetically
just that David's beloved BBN theory has the effect of lowering globular
cluster ages, as he would have liked.

One may wonder what further dramatic improvements might be possible in
advance of improved distance determinations.  One area where significant
improvement is possible is in the determination of the age of individual
globular clusters.  Here, the MSTO method presently leads to uncertainties
on the order of $\pm 1-2$ Gyr.  There is, however, nothing sacred about
the use of the MSTO in the analysis.  Indeed, as we demonstrated in a
separate analysis \cite{chab25}, the theoretical uncertainties remain
comparable, but observational  uncertainties are greatly reduced if one
considers determining ages using a point on the subgiant branch that is
0.05 mag redder than the turnoff point .  At the time, computational
limits prohibited carrying out detailed Monte Carlo estimates further up
the subgiant branch.  However, the speed of computers has altered this
situation.  We are currently undertaking a new analysis which will
systematically explore using a variety of points along the CM diagram for
globular clusters to explore globular cluster ages.  We expect in this
way to be able to reduce the individual  uncertainty in relative age of
globular clusters by up to an order of  magnitude.  This will in turn
lead to a more restricted distribution in  mean ages of the oldest
globular clusters.

In conclusion, the past 5 years has seen dramatic improvements in determining
the age of globular clusters.  The direction has been precisely where
David Schramm imagined.  The best fit ages are currently 2-3 Gyr smaller than
they were a decade ago.  At the same time, with improved observational
limits on the Hubble constant, and with tentative support for an 
accelerating Universe, the importance of further improving our knowledge
of the age of globular clusters in order to constrain cosmology is, if 
anything, stronger than it was before.  If David were still with us, he
would be smiling.

\section*{Acknowledgment}   
I thank my collaborators, Brian Chaboyer, Pierre Demarque, and Peter
Kernan for their contributions to the results described here.   This
research is supported by the DOE and by funds from Case Western Reserve
University.

\end{document}